\documentclass[iop]{emulateapj}  

\usepackage{color}
\usepackage{hyperref}  
\usepackage{breakurl}  

\newcommand\kms{\ifmmode{\rm km\thinspace s^{-1}}\else km\thinspace s$^{-1}$\fi}
\newcommand\epic{EPIC~219511354}   
\newcommand\epicthird{EPIC~219552514}
\newcommand\epicsecond{EPIC~219568666}
\newcommand\epicfirst{EPIC~219394517}
\newcommand\rup{Ruprecht~147}

\shortauthors{Torres et al.}
\shorttitle{Eclipsing binary in \rup}

\begin{document} 
\submitted{Accepted for publication in The Astrophysical Journal}

\title{Eclipsing binaries in the open cluster Ruprecht 147. IV: The
  active triple system \epic}

\author{
Guillermo Torres\altaffilmark{1},
Andrew Vanderburg\altaffilmark{2,3},
Jason L.\ Curtis\altaffilmark{4}, 
Adam L.\ Kraus\altaffilmark{5}, and
Eric Gaidos\altaffilmark{6}
}

\altaffiltext{1}{Center for Astrophysics $\vert$ Harvard \&
  Smithsonian, 60 Garden St., Cambridge, MA 02138, USA;
  gtorres@cfa.harvard.edu}

\altaffiltext{2}{Department of Physics and Kavli Institute for Astrophysics and Space Research, Massachusetts Institute of Technology, Cambridge, MA 02139, USA}

\altaffiltext{3}{Department of Astronomy, University of Wisconsin-Madison, Madison, WI 53706, USA}

\altaffiltext{4}{American Museum of Natural History, Central Park
  West, New York, NY, USA}
  
\altaffiltext{5}{Department of Astronomy, The University of Texas at
  Austin, Austin, TX 78712, USA}

\altaffiltext{6}{Department of Earth Sciences, University of Hawai'i at M\"{a}noa, Honolulu, HI 96822, USA}


\begin{abstract} 

We report follow-up spectroscopic observations of the 1.62~day,
K-type, detached, active, near-circular, double-lined eclipsing binary
\epic\ in the open cluster \rup, identified previously on the basis of
photometric observations from the Kepler/K2 mission. This is the
fourth eclipsing system analyzed in this cluster. A combined analysis
of the light curve and radial velocities yields accurate masses of
$M_{\rm Aa} = 0.912 \pm 0.013~M_{\sun}$ and $M_{\rm Ab} = 0.822 \pm
0.010~M_{\sun}$ for the primary (star Aa) and secondary (Ab), along
with radii of $R_{\rm Aa} = 0.920 \pm 0.016~R_{\sun}$ and $R_{\rm Ab}
= 0.851 \pm 0.016~R_{\sun}$, and effective temperatures of $5035 \pm
150$ and $4690 \pm 130$~K, respectively. Comparison with current
models of stellar evolution for the known age and metallicity of the
cluster reveals that both radii are larger (by 10--14\%) and both temperatures
cooler (by $\sim$6\%) than theoretically predicted, as is often seen in M dwarfs.
This is likely caused by the significant stellar activity in the
system, manifested here by 6\% peak-to-peak out-of-eclipse
variability, a filled-in H$\alpha$ line, and its detection as an X-ray
source. We also find \epic\ to be a hierarchical triple system, with a
low-mass tertiary in an eccentric 220~day orbit.

\end{abstract}

\section{Introduction}
\label{sec:introduction}

\rup\ (NGC~6674) is a middle-aged open cluster with slightly
supersolar metallicity \citep[${\rm [Fe/H]} = +0.10$;
  e.g.,][]{Curtis:2018}, which was observed in late 2015 by NASA's K2
mission (Campaign~7).  It is remarkable in that it has no less than
five eclipsing systems identified by \cite{Curtis:2016} that are
detached, relatively bright, and therefore quite amenable to
follow-up. In previous papers in this series we have presented the
analysis of three of them: \epicfirst, \epicsecond, and
\epicthird\ \citep[][hereafter Papers~I, II, and III]{Torres:2018,
  Torres:2019, Torres:2020}. These yielded accurate determinations of
the masses, radii, and effective temperatures for the binary
components that are in good agreement with stellar evolution theory.
Importantly, these three systems also provided an accurate weighted mean
age for \rup\ of $2.67^{+0.39}_{-0.55}$~Gyr. When combined with
age determinations in other populations, \rup\ can then serve as a benchmark for
studying the evolution of rotation in Sun-like stars as they spin down
via magnetic braking \citep[see][]{Curtis:2020}, and even as a benchmark
for the white dwarf initial-final mass relation \citep[e.g.,][]{Marigo:2020}.

This paper deals with a fourth eclipsing system in \rup, \epic\ (also
known as Gaia DR2~4087847862581712768), which is a pair of detached
K-type main-sequence stars with an orbital period of 1.62~days.
Results from the Gaia mission confirm its membership in the
cluster. Our original goal was to analyze the K2 photometry and new
spectroscopic observations we have obtained, in order to provide a
further check of stellar evolution theory in a cluster with known
metallicity, and to add an independent age determination. Instead, we
have found that the system is unsuitable for that purpose due to
significant systematic discrepancies with the models, but provides
instead a particularly interesting example of the effects of stellar
activity on the global properties of stars with convective envelopes.

We begin by presenting the photometric, spectroscopic, and imaging
observations in Section~\ref{sec:observations}, where we also report
the discovery that \epic\ is in fact a hierarchical triple system with
an unseen low-mass outer companion.  Our lightcurve analysis is
presented in Section~\ref{sec:analysis}, followed by a determination
of the absolute dimensions (masses, radii, temperatures, etc.)\ of the
binary components in Section~\ref{sec:dimensions}. We discuss the
rotation of the components and various indicators of stellar activity
in Section~\ref{sec:rotation}. Then in Section~\ref{sec:models} we
present a comparison of the measured properties for \epic\ against
current models of stellar evolution, where we highlight the
disagreements with theory stemming from the activity. A discussion of
this and other issues, along with our conclusions, may be found in
Section~\ref{sec:conclusions}.

\section{Observations}
\label{sec:observations}

\subsection{Photometry}
\label{sec:photometry}

\epic\ was observed by the Kepler spacecraft during Campaign~7 of its extended K2 mission,\footnote{Data release notes are available online at \url{https://archive.stsci.edu/missions/k2/doc/drn/KSCI-19125-002_K2-DRN9_C7.pdf}.} for a total of 83 days between 2015 October 4 and 2015 December 26. Onboard processing included co-adding the data as they were collected, to form ``long-cadence'' observations with a 29.4 minute integration time. At end of the campaign, the data were downlinked to Earth, calibrated with pipeline routines run at the NASA Ames Research Center, and released to the public. We downloaded the pixel-level postage stamp data for \epic\ from the Mikulski Archive for Space Telescopes (MAST),\footnote{\url{http://archive.stsci.edu/}} and extracted lightcurves using a strategy similar to that employed in Papers~I--III. In particular, we extracted lightcurves using a circular moving aperture with a 6\arcsec\ radius (excluding the first day and a half when the spacecraft was thermally settling), and removed systematic artifacts caused by the K2 mission's unstable pointing using the methods described by \citet{Vanderburg:2014} and \citet{Vanderburg:2016}. After performing a first-pass systematics correction, we then refined the corrections by simultaneously fitting the shapes of the binary eclipses, the spacecraft systematics, and the intrinsic stellar variability, as described by \citet{Vanderburg:2016}. As shown later, out-of-eclipse stellar variability is quite obvious, and
was removed for the detailed analysis below by fitting a spline function with the eclipses masked out. We then normalized the photometry by dividing this fit into the systematics-corrected data.
The resulting lightcurve is shown during the eclipses in Section~\ref{sec:analysis}, and out of eclipse in Section~\ref{sec:rotation}, and is available for download in Table \ref{tab:photometry} with and without the normalization. Fifty pairs of eclipses (primary and secondary) are included within the 83 day duration of the observations.

%
\setlength{\tabcolsep}{6pt}  
\begin{deluxetable}{ccc}
\tablewidth{1.0\columnwidth}
\tablecaption{Detrended K2 Photometry of \epic\ \label{tab:photometry}}
\tablehead{
\colhead{BJD} &
\colhead{Residual flux} &
\colhead{Residual flux}
\\
\colhead{(2,400,000+)} &
\colhead{(flattened)} &
\colhead{(not flattened)}
}
\startdata
  57301.4866 &  0.99381119 &  0.99990281 \\
  57301.5070 &  0.99446314 &  0.99993898 \\
  57301.5275 &  0.99551235 &  1.00029860 \\
  57301.5479 &  0.99633834 &  1.00039691 \\
  57301.5683 &  0.99653720 &  0.99985861 
\enddata

\tablecomments{K2 photometry after removal of instrumental effects and
  long-term drifts, and normalized by dividing by a spline fit to the
  out-of-eclipse observations (second column). The third column
  contains the same data without the normalization. (This table is
  available in its entirety in machine-readable form.)}

\end{deluxetable}
\setlength{\tabcolsep}{6pt}  

\subsection{Spectroscopy}
\label{sec:spectroscopy}

Spectroscopic observations of \epic\ were carried out at the Center
for Astrophysics between 2015 October and 2021 June, using the
Tillinghast Reflector Echelle Spectrograph
\citep[TRES;][]{Szentgyorgyi:2007, Furesz:2008}. This is a fiber-fed,
bench-mounted instrument that is attached to the 1.5m Tillinghast
reflector at the Fred L.\ Whipple Observatory on Mount Hopkins
(Arizona, USA).  We collected a total of 19 usable spectra at a
resolving power of $R \approx 44,000$, which cover the wavelength
region 3800--9100~\AA\ in 51 orders. The signal-to-noise ratios in the
order centered at $\sim$5187~\AA\ containing the \ion{Mg}{1}~b triplet
range from 15 to 25 per resolution element of 6.8~\kms.

All our spectra display double lines. Radial velocities were measured
in the order centered on the \ion{Mg}{1}~b triplet using the
two-dimensional cross-correlation algorithm {\tt TODCOR\/}
\citep{Zucker:1994}.  Templates for each star were selected from a
pre-computed library of synthetic spectra that are based on model
atmospheres by R.\ L.\ Kurucz, and a line list tuned to better match
the spectra of real stars \citep[see][]{Nordstrom:1994,
  Latham:2002}. These templates cover a limited wavelength region of
$\sim$300~\AA\ centered around 5187~\AA, and are parameterized in
terms of the effective temperature ($T_{\rm eff}$), surface gravity
($\log g$), rotational broadening ($v \sin i$ when seen in projection)
and metallicity ([Fe/H]). We adopted $\log g$ values of 4.5 for both
stars, very near our final values from the analysis below. The
metallicity was set to solar. This is close to the actual value
measured for \rup, which is ${\rm [Fe/H]} = +0.10$ \citep{Curtis:2018,
  Bragaglia:2018}, and the minor difference has no impact on the
velocities. We optimized the remaining parameters for the templates in
the manner described by \cite{Torres:2002}, based on grids of
cross-correlations over wide ranges in $T_{\rm eff}$ and $v \sin
i$. While our template library has relatively course steps in
temperature and $v \sin i$, we derived estimates of the actual
temperatures and rotational velocities of the components by
interpolation between grid points, obtaining $5310 \pm 200$~K and
$5100 \pm 200$~K for the primary and secondary, along with $v \sin i$
values of $32 \pm 3$ and $31 \pm 4~\kms$, respectively. The
uncertainties are based on the scatter from the individual spectra,
conservatively increased to account for possible systematic errors
in the models by adding 100~K and 2~\kms\ in quadrature.  The
radial velocity determinations were made using the nearest templates
in our grid, with 5250~K and 5000~K for the temperatures, and
projected rotational velocities of 30~\kms\ for both stars.

In a few of the spectra we noticed varying levels of contamination by
moonlight. We removed this effect from the velocities by rederiving
them with {\tt TRICOR}, which is an extension of {\tt TODCOR\/} to
three dimensions \citep{Zucker:1995}. For the third template we
adopted parameters appropriate for the Sun.  The final velocities are
listed in Table~\ref{tab:rvs}. The flux ratio we obtained between the
fainter secondary and the primary star is $0.53 \pm 0.02$,
corresponding to the mean wavelength of our observations (5187~\AA).

\setlength{\tabcolsep}{3.5pt}  
\begin{deluxetable}{c@{}c@{}c@{}cc}
\tablewidth{1.0\columnwidth}
\tablecaption{Heliocentric Radial-velocity Measurements of \epic \label{tab:rvs}}
\tablehead{
\colhead{HJD} &
\colhead{$RV_{\rm Aa}$} &
\colhead{$RV_{\rm Ab}$} &
\colhead{Inner} &
\colhead{Outer}
\\
\colhead{(2,400,000$+$)} &
\colhead{(\kms)} &
\colhead{(\kms)} &
\colhead{Phase} &
\colhead{Phase}
}
\startdata
 57297.6914 &   $ 80.20 \pm 1.26$\phn      &   $ 11.10 \pm 2.03$\phn      &    0.9483  &  0.9307 \\
 57648.6636 &   $-56.80 \pm 1.34$\phn\phs  &   $140.50 \pm 2.16$\phn\phn  &    0.3233  &  0.5228 \\
 57853.9931 &   $ 92.80 \pm 1.82$\phn      &   $-20.20 \pm 2.94$\phn\phs  &    0.9092  &  0.4543 \\
 57865.9773 &   $-60.10 \pm 1.35$\phn\phs  &   $146.50 \pm 2.17$\phn\phn  &    0.2975  &  0.5087 \\
 57885.9165 &   $ 94.30 \pm 1.65$\phn      &   $-21.00 \pm 2.65$\phn\phs  &    0.5901  &  0.5991 \\
 57886.9386 &   $-62.10 \pm 1.57$\phn\phs  &   $150.60 \pm 2.53$\phn\phn  &    0.2202  &  0.6037 \\
 57907.9425 &   $-52.00 \pm 1.94$\phn\phs  &   $139.20 \pm 3.12$\phn\phn  &    0.1691  &  0.6990 \\
 57933.9168 &   $-53.30 \pm 1.77$\phn\phs  &   $145.80 \pm 2.85$\phn\phn  &    0.1823  &  0.8169 \\
 58004.7189 &   $137.20 \pm 1.59$\phn\phn  &   $-56.30 \pm 2.57$\phn\phs  &    0.8320  &  0.1380 \\
 58033.6031 &   $121.80 \pm 1.48$\phn\phn  &   $-47.10 \pm 2.38$\phn\phs  &    0.6391  &  0.2691 \\
 58037.6155 &   $-24.31 \pm 1.65$\phn\phs  &   $116.73 \pm 2.65$\phn\phn  &    0.1128  &  0.2873 \\
 58281.8504 &   $133.60 \pm 1.42$\phn\phn  &   $-65.80 \pm 2.28$\phn\phs  &    0.6840  &  0.3952 \\
 58300.8963 &   $ -4.70 \pm 1.36$\phs      &   $ 86.80 \pm 2.19$\phn      &    0.4259  &  0.4816 \\
 58633.9426 &   $151.50 \pm 1.78$\phn\phn  &   $-62.70 \pm 2.87$\phn\phs  &    0.7495  &  0.9924 \\
 58662.8934 &   $110.89 \pm 1.98$\phn\phn  &   $-25.18 \pm 3.18$\phn\phs  &    0.5977  &  0.1238 \\
 59143.5896 &   $ 72.30 \pm 2.15$\phn      &   $  7.90 \pm 3.46$          &    0.9478  &  0.3044 \\
 59341.9628 &   $-58.90 \pm 1.69$\phn\phs  &   $157.40 \pm 2.72$\phn\phn  &    0.2452  &  0.2043 \\
 59371.8955 &   $136.40 \pm 1.65$\phn\phn  &   $-67.50 \pm 2.65$\phn\phs  &    0.6987  &  0.3401 \\
 59384.8673 &   $137.00 \pm 2.00$\phn\phn  &   $-72.20 \pm 3.21$\phn\phs  &    0.6959  &  0.3989 
\enddata

\tablecomments{Orbital phases for the inner orbit (stars Aa and Ab)
  are counted from the reference time of primary eclipse $T_0$ given in
  Table~\ref{tab:mcmc}, and those for the outer orbit from the time of
  periastron passage $T_{\rm AB}$ in Table~\ref{tab:specorbit}.}

\end{deluxetable}
\setlength{\tabcolsep}{6pt}  

\begin{figure}
\epsscale{1.15}
\plotone{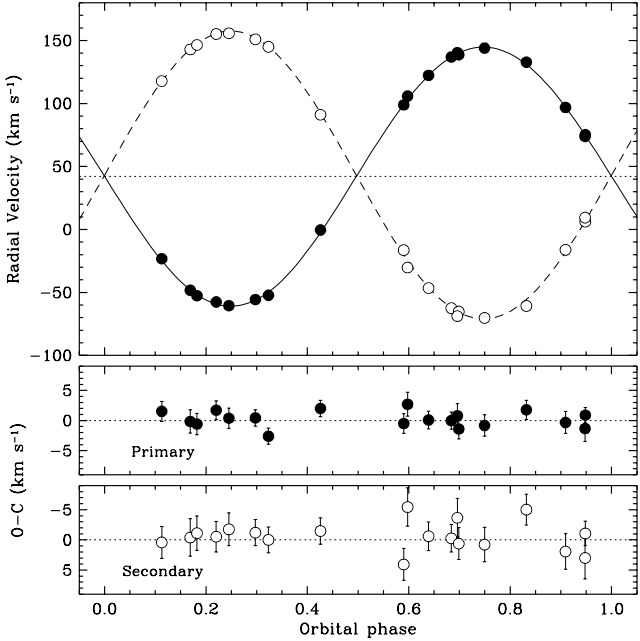}

\figcaption{Radial-velocity measurements for \epic, with our adopted
  model for the inner orbit (Table~\ref{tab:specorbit}). Primary and
  secondary measurements are represented with filled and open circles,
  respectively, and have the motion in the outer orbit removed. The
  dotted line marks the center-of-mass velocity of the triple system.
  Error bars are smaller than the symbol size in the top panel, but
  are seen in the lower panels, which display the residuals. Phases
  are counted from the reference time of primary eclipse
  (Table~\ref{tab:mcmc}).\label{fig:inner}}

\end{figure}

\begin{figure}
\epsscale{1.15}
\plotone{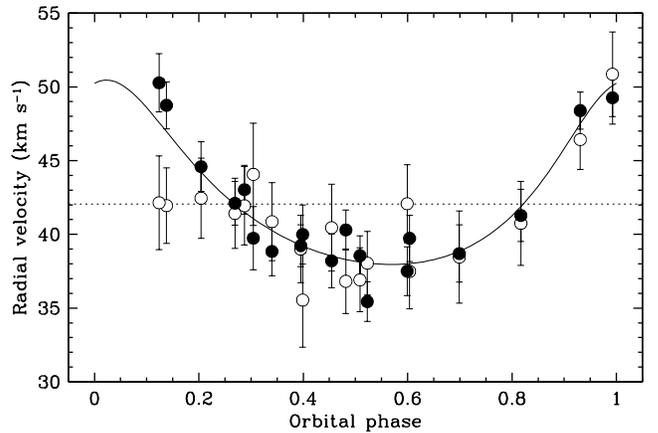}

\figcaption{Radial-velocity measurements and model for \epic\ in the
  outer orbit, with the motion in the inner orbit removed from the
  measurements. Symbols and the dotted line are as in
  Figure~\ref{fig:inner}. \label{fig:outer}}

\end{figure}

A spectroscopic orbital solution based on these velocities gave rms
residuals near 5~\kms\ for both stars, which are larger than we
expected from spectra of the quality we have.  Examination of the
residuals versus time revealed deviations from zero with a very
similar pattern for the primary and secondary, strongly suggesting the
presence of an unseen third star in the system. A triple-star solution
assuming independent inner and outer Keplerian orbits indicates the
outer orbit has a period of about 220 days, and a modest eccentricity
of $e \approx 0.27$. Our observations cover 9.5 cycles of the outer
orbit. The inner orbit, in turn, shows an eccentricity that is not
statistically significant, given the data at hand. Hereafter we refer
to the stars in the inner binary as Aa and Ab, and to the third star
as B.  We report the elements of the inner and outer orbits in
Table~\ref{tab:specorbit}, and present visual representations of the
observations with our best-fit model in Figures~\ref{fig:inner} and
\ref{fig:outer}.  The inner period in this solution has been fixed to
the value from the lightcurve solution described later.  The rms
residuals in this fit are reduced by a factor of two compared to our
initial solution.

\setlength{\tabcolsep}{10pt}
\begin{deluxetable}{lc}
\tablewidth{1.0\columnwidth}
\tablecaption{Spectroscopic Orbital Solution for \epic \label{tab:specorbit}}
\tablehead{ \colhead{~~~~~~~~~~~Parameter~~~~~~~~~~~} & \colhead{Value}}
\startdata
 $P_{\rm A}$ (days)                 &  1.6220554 (fixed)                     \\ [1ex]
 $T_0$ (HJD$-$2,400,000)            &  $58288.8512 \pm 0.0026$               \\ [1ex]
 $\gamma$ (\kms)                    &  $+42.57 \pm 0.44$                     \\ [1ex]
 $K_{\rm Aa}$ (\kms)                &  $102.89 \pm 0.57$                     \\ [1ex]
 $K_{\rm Ab}$ (\kms)                &  $114.28 \pm 0.81$                     \\ [1ex]
 $e_{\rm A}$ (\kms)                 &  $0.0068 \pm 0.0050$                   \\ [1ex]
 $\omega_{\rm Aa}$ (degree)         &  $204 \pm 34$                          \\ [1ex]
 $P_{\rm AB}$ (days)                &  $220.4 \pm 1.2$                       \\ [1ex]
 $T_{\rm AB}$ (HJD$-$2,400,000)     &  $57974 \pm 11$                        \\ [1ex]
 $K_{\rm A}$ (\kms)                 &  $6.25 \pm 0.81$                       \\ [1ex]
 $e_{\rm AB}$ (\kms)                &  $0.27 \pm 0.11$                       \\ [1ex]
 $\omega_{\rm A}$ (degree)          &  $345 \pm 21$                          \\ [1ex]
\noalign{\hrule} \\ [-1.5ex]
\multicolumn{2}{c}{Derived quantities} \\ [1ex]
\noalign{\hrule} \\ [-1.5ex]
 $M_{\rm Aa} \sin^3 i$ ($M_{\sun}$) &  $0.906 \pm 0.013$                     \\ [1ex]
 $M_{\rm Ab} \sin^3 i$ ($M_{\sun}$) &  $0.816 \pm 0.010$                     \\ [1ex]
 $q \equiv M_{\rm Ab}/M_{\rm Aa}$   &  $0.9003 \pm 0.0087$                   \\ [1ex]
 $a_{\rm Aa} \sin i$ ($10^6$ km)    &  $2.295 \pm 0.013$                     \\ [1ex]
 $a_{\rm Ab} \sin i$ ($10^6$ km)    &  $2.549 \pm 0.018$                     \\ [1ex]
 $a_{\rm A} \sin i$ ($R_{\sun}$)    &  $6.963 \pm 0.029$                     \\ [1ex]
 $a_{\rm AB} \sin i$ ($10^6$ km)    &  $18.2 \pm 2.1$                        \\ [1ex]
 $f(M)$ ($M_{\sun}$)                &  $0.0050 \pm 0.0017$                   \\ [1ex]
 $M_{\rm B} \sin i / (M_{\rm AB}+M_{\rm B})^{2/3}$ ($M_{\sun}$) &  $0.171 \pm 0.019$      \\ [1ex]
\noalign{\hrule} \\ [-1.5ex]
\multicolumn{2}{c}{Other quantities pertaining to the fit} \\ [1ex]
\noalign{\hrule} \\ [-1.5ex]
 $\sigma_{\rm Aa}$, $\sigma_{\rm Ab}$ (\kms)       &  1.62,  2.61            \\ [1ex]
 $N_{\rm Aa}$,  $N_{\rm Ab}$        &  19, 19                                \\ [1ex]
 Time span (days)                   &  2087.2                                \\ [1ex]
 Cycles covered in inner orbit      &  1286.7                                \\ [1ex]
 Cycles covered in outer orbit      &  9.5 
\enddata

\tablecomments{The symbol $T_0$ represents a reference time of eclipse
  for the inner binary; $T_{\rm AB}$ is a reference time of periastron
  passage in the outer orbit; $a_{\rm AB}$ is the semimajor axis of
  the eclipsing binary in the outer orbit; $f(M)$ is the mass function
  of the outer binary.}

\end{deluxetable}
\setlength{\tabcolsep}{6pt}

We searched for the lines of the third star in our spectra using {\tt
  TRICOR}, for observations uncontaminated by moonlight. For
contaminated spectra we used {\tt QUADCOR}, which is a further
extension of {\tt TODCOR\/} to four dimensions \citep{Torres:2007}. No
compelling sign of the tertiary was found, which we estimate we would
have seen if it were any brighter than about 1\% of the light of the
primary star.

\subsection{Imaging}
\label{sec:imaging}

\begin{figure}
\epsscale{1.15}
\plotone{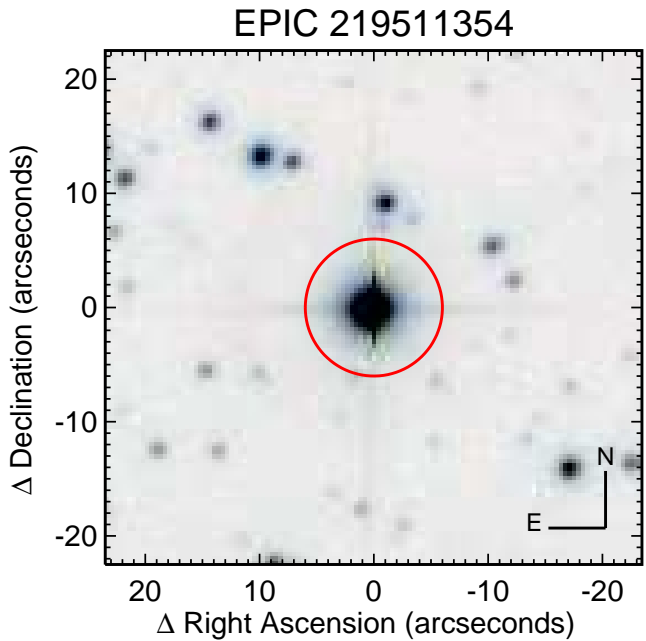}
\figcaption{CFHT $r$-band image of the field of \epic. The circle
  represents the 6\arcsec\ photometric aperture used to extract the K2
  photometry. \label{fig:image}}
\end{figure}

Compared to a typical star in the field, membership of \epic\ in a
cluster such as \rup\ carries an increased risk that objects in the
vicinity may fall in the photometric aperture and affect the light
curve. This does not appear to be the case here.
Figure~\ref{fig:image} shows a seeing-limited image in a bandpass
similar to Sloan $r$ (close to Kepler's $K\!p$ bandpass), taken in
2008 by \cite{Curtis:2013} with the MegaCam instrument
\citep{Hora:1994} on the Canada-France-Hawaii-Telescope (CFHT). Many
of the stars in this image are also listed in the Gaia EDR3 catalog
\citep{Gaia:2021}, but none are inside the 6\arcsec\ radius aperture
and are bright enough to have an impact.

In order to search for companions closer than seeing-limited imaging
can reveal, we observed \epic\ at the Keck~II Observatory with the NIRC2 infrared adaptive optics imager on UT 2016 May 12, to obtain Natural Guide Star adaptive optics imaging. We used the $K'$ filter and followed the standard observing strategy described by \cite{Kraus:2016} and previously reported for Ruprecht 147 targets by \cite{Torres:2018, Torres:2019}. We obtained a short sequence of five images in vertical angle model, which stops the telescope rotator in order to allow the sky to rotate overhead and to create a uniform stellar point spread function (PSF) that enhances subsequence PSF calibration. Each image had a total integration time of 10~sec, and to avoid saturation, each image was divided into 10 exposures of 1~sec each that were coadded on the detector. The integration time allowed for four Fowler samples in each individual 1~sec exposure.

The images were analyzed using the methods described by \cite{Kraus:2016}. To briefly summarize, for each image the PSF of the primary star was subtracted using two methods: an azimuthal-median profile, and an empirical PSF calibrator that most closely matches the science target, chosen from among the 1000 most contemporaneous images of other stars that appeared to be single. Within each residual image, the flux as a function of position was measured in apertures of radius 40~mas, centered on each pixel, and the noise was estimated from the RMS of fluxes within concentric rings around the primary star. Finally, the detections and detection limits were estimated from the strehl-weighted sum of the detection significances in the image stack, and any location with a total significance of $>6\sigma$ was visually inspected to determine if it was a residual speckle or cosmic ray. 

We ultimately determined that there are two faint sources that share the NIRC2 field of view with \epic, one close to the star ($\rho = 1303.1 \pm 1.9$~mas, $\theta = 189\fdg13 \pm 0\fdg8$, $\Delta K' = 7.024 \pm 0.024$~mag), and one much farther away ($\rho = 8383 \pm 5$~mas, $\theta = 156\fdg17 \pm 0\fdg03$, $\Delta K' = 8.30 \pm 0.08$~mag) that is outside the 6\arcsec\ photometric aperture. With only a single epoch, we are unable to determine if either is comoving, but given their faintness, they should not significantly bias the fitting of the light curves. The corresponding limits on additional companions were $\Delta K' = 5.5$~mag at $\rho = 150$~mas, $\Delta K' = 8.1$~mag at $\rho = 500$~mas, and $\Delta K' = 8.8$~mag at $\rho > 1\farcs5$.

\section{Lightcurve Analysis}
\label{sec:analysis}

The Kepler/K2 light curve of \epic\ was analyzed in the same way as
previous papers in this series, using the Nelson-Davis-Etzel binary
model \citep{Etzel:1981, Popper:1981}, which is appropriate for
well-detached systems such as this in which the stars are essentially
spherical (see below). Specifically, we used the implementation of
this model in the {\tt eb\/} code of \cite{Irwin:2011}, which
facilitates its use within a Markov chain Monte Carlo (MCMC)
environment.

The adjustable parameters we considered are the orbital period of the
inner binary ($P_{\rm A}$), a reference time of primary eclipse
($T_0$), the central surface brightness ratio in the Kepler band ($J
\equiv J_{\rm Ab}/J_{\rm Aa}$), the sum of the relative radii
normalized by the semimajor axis ($r_{\rm Aa}+r_{\rm Ab}$) and their
ratio ($k \equiv r_{\rm Ab}/r_{\rm Aa}$), the cosine of the
inclination angle ($\cos i$), the eccentricity parameters
$\sqrt{e_{\rm A}} \cos\omega_{\rm Aa}$ and $\sqrt{e_{\rm A}}
\sin\omega_{\rm Aa}$, where $e_{\rm A}$ is the inner binary
eccentricity and $\omega_{\rm Aa}$ the argument of periastron for the
primary, and an out-of-eclipse brightness level in magnitude units
($m_0$). We adopted a linear limb-darkening law with fixed
coefficients for the Kepler band taken from \cite{Claret:2011}, as
tests with a quadratic law gave no improvement and did not change the
geometric parameters. The coefficients were interpolated in the
tabulation from those authors based on the effective temperatures and
surface gravities for the components described in
Section~\ref{sec:dimensions} below, and the metallicity of ${\rm
  [Fe/H]} = +0.10$ for \rup. The coefficients are 0.681 for the
primary and 0.717 for the secondary. As a final free parameter we
included a multiplicative scale factor for the observational errors,
which we assumed initially to have the same (arbitrary) value of 0.002
mag for all epochs.

Because the normalization of the light curve described in
Section~\ref{sec:photometry} artificially removes any variations out
of eclipse, gravity darkening and reflection become irrelevant. For
consistency, we therefore used the option in the {\tt eb\/} code that
suppresses those effects in calculating the binary model. As the
out-of-eclipse portions of the light curve then provide no additional
information, we retained for the analysis only sections within 0.12 in
phase units from each eclipse, equivalent to about one and a half
times the total eclipse duration. Third light was set to zero given
the absence of any nearby companions within the photometric aperture,
and the non-detection of the tertiary in our spectra. Tests with third
light set to the maximum allowed by the spectroscopy resulted in only
very small changes in some of the geometric parameters that are well
within our final uncertainties described below.

In eclipsing binaries such as \epic\ with partial eclipses and similar
components, the radius ratio $k$ is often strongly correlated with
other properties such as $J$, and as a result it can be poorly
constrained by the photometry. This problem can be alleviated by
requiring the solution to be consistent with an independently measured
light ratio, such as from spectroscopy \citep[see,
  e.g.,][]{Andersen:1980}. This is effective because the light ratio
depends very strongly on the radius ratio: $\ell_{\rm Ab}/\ell_{\rm
  Aa} \propto k^2$. For this we used our measurement from
Section~\ref{sec:spectroscopy}, which we transformed to the Kepler
band using synthetic spectra based on PHOENIX models from
\cite{Husser:2013}, for the properties of the components reported
below. We applied this constraint ($\ell_{\rm Ab}/\ell_{\rm Aa} = 0.62
\pm 0.03$) in the form of a Gaussian prior.

The cadence of the Kepler observations (29.4 minutes) is a
non-negligible fraction of the orbital period of the inner binary
(1.62 days), equivalent to almost 0.013 in phase units. In order to
avoid biases from smearing, we oversampled the model light curve at
each iteration of our solution, and then integrated over the 29.4
minute duration of each cadence prior to the comparison with the
observations \citep[see][]{Gilliland:2010, Kipping:2010}. In
principle, an additional bias may come from the finite light travel
time of the binary in the outer orbit, advancing or delaying the times
of eclipse. However, the effect is very small in this case, ranging
from +0.5 minutes $-1.0$ minutes over the 81 days of photometric
coverage. Nevertheless, we chose to apply these corrections to the
individual times of observation, which we derived from our
spectroscopic orbital solution.

\begin{figure*}
\epsscale{1.15}
\hskip 5pt \includegraphics[height=17.6cm, angle=270]{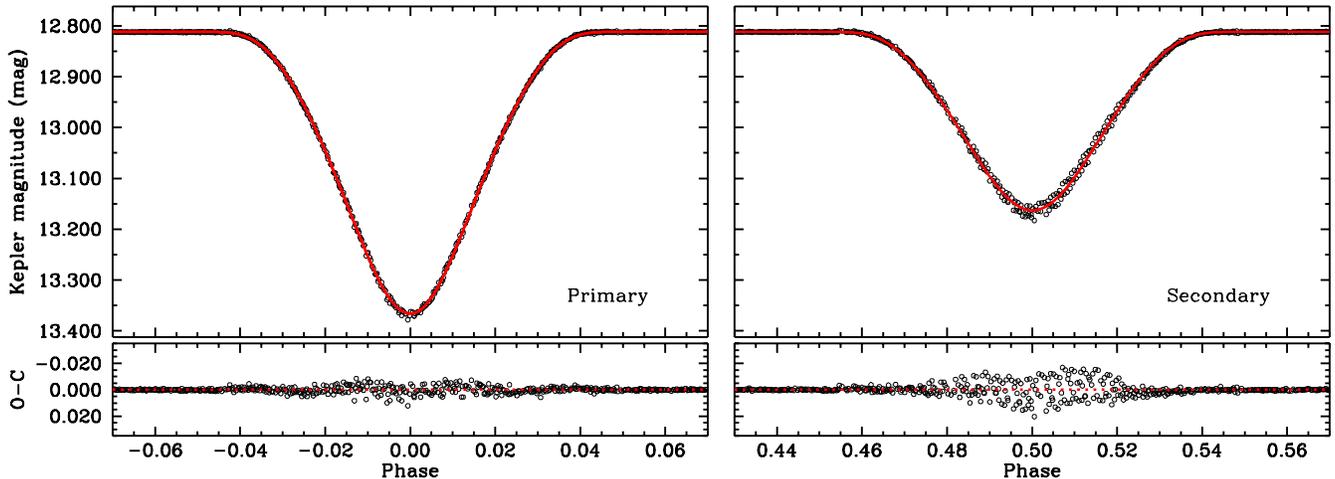}

\figcaption{K2 observations and model light curve for \epic\ during primary and secondary eclipse.\label{fig:mcmc}}

\end{figure*}

We carried out our lightcurve analysis using the {\tt
  emcee\/}\footnote{\url{https://github.com/dfm/emcee}} code of
\cite{Foreman-Mackey:2013}, which is a Python implementation of the
affine-invariant MCMC ensemble sampler proposed by
\cite{Goodman:2010}. We used 100 walkers with chain lengths of 5,000
each, after discarding the burn-in. All parameters used uniform or
log-uniform priors over suitable ranges listed in
Table~\ref{tab:mcmc}. We verified convergence by visual examination of
the chains, and by requiring a Gelman-Rubin statistic of 1.05 or
smaller for each parameter \citep{Gelman:1992}. The results are
reported in Table~\ref{tab:mcmc}. The posterior distributions of the
derived quantities in the bottom section of the table were constructed
directly from the MCMC chains of the adjustable parameters involved.
The eccentricity of the orbit is formally significant at about the
4.3$\sigma$ level.  The oblateness of the stars, calculated as
prescribed by \cite{Binnendijk:1960}, is 0.003, which is well below
the upper limit of 0.04 considered safe for the Nelson-Davis-Etzel
binary model \citep[see, e.g.,][]{Popper:1981}, justifying its use.
Figure~\ref{fig:mcmc} shows the K2 observations together with our
model.

%
\setlength{\tabcolsep}{5pt}
\begin{deluxetable}{lcc}
\tablewidth{1.0\columnwidth}
\tablecaption{Results from our MCMC Lightcurve Analysis of \epic \label{tab:mcmc}}
\tablehead{ \colhead{~~~~~~~Parameter~~~~~~~} & \colhead{Value} & \colhead{Prior} }
\startdata
 $P_{\rm A}$ (days)               &  $1.6220554^{+0.0000011}_{-0.0000011}$    & [1.6, 1.7] \\ [1ex]
 $T_0$ (HJD$-$2,400,000)          &  $57336.706796^{+0.000019}_{-0.000019}$      & [57336, 57337] \\ [1ex]
 $J$                              &  $0.70167^{+0.00079}_{-0.00073}$          & [0.4, 1.0] \\ [1ex]
 $r_{\rm Aa}+r_{\rm Ab}$          &  $0.25377^{+0.00021}_{-0.00021}$          & [0.01, 0.50] \\ [1ex]
 $k \equiv r_{\rm Ab}/r_{\rm Aa}$ &  $0.924^{+0.013}_{-0.013}$                & [0.5, 1.5] \\ [1ex]
 $\cos i$                         &  $0.06665^{+0.00062}_{-0.00067}$          & [0, 1] \\ [1ex]
 $\sqrt{e_{\rm A}} \cos\omega_{\rm Aa}$  &  $+0.00213^{+0.00059}_{-0.00051}$  & [$-$1, 1] \\ [1ex]
 $\sqrt{e_{\rm A}} \sin\omega_{\rm Aa}$  &  $-0.0682^{+0.0082}_{-0.0075}$     & [$-$1, 1] \\ [1ex]
 $m_0$ (mag)                      &  $12.811994^{+0.000082}_{-0.000081}$      & [12, 13] \\ [1ex]
 $f_{\rm K2}$                     &  $1.439^{+0.024}_{-0.024}$                & [$-$5, 5] \\ [1ex]
\noalign{\hrule} \\ [-1.5ex]
\multicolumn{3}{c}{Derived quantities} \\ [1ex]
\noalign{\hrule} \\ [-1.5ex]
 $r_{\rm Aa}$                     &  $0.13190^{+0.00082}_{-0.00082}$          & \nodata \\ [1ex]
 $r_{\rm Ab}$                     &  $0.12187^{+0.00093}_{-0.00092}$          & \nodata \\ [1ex]
 $i$ (degree)                     &  $86.178^{+0.038}_{-0.036}$               & \nodata \\ [1ex]
 $e_{\rm A}$                      &  $0.0047^{+0.0011}_{-0.0011}$             & \nodata \\ [1ex]
 $\omega_{\rm Aa}$ (degree)       &  $271.78^{+0.68}_{-0.49}$                 & \nodata \\ [1ex]
 $\ell_{\rm Ab}/\ell_{\rm Aa}$    &  $0.590^{+0.016}_{-0.016}$                & $G(0.62, 0.03)$ \\ [1ex]
 $J_{\rm ave}$                    &  $0.69102^{+0.00078}_{-0.00072}$          & \nodata 
\enddata

\tablecomments{The values listed correspond to the mode of the
  respective posterior distributions, and the uncertainties represent
  the 68.3\% credible intervals. Priors in square brackets are uniform
  over the specified ranges, except the one for $f_{\rm K2}$, which is
  log-uniform, and the one for the light ratio, which is Gaussian,
  indicated above as $G({\rm mean}, \sigma)$.}

\end{deluxetable}
\setlength{\tabcolsep}{6pt}

There is significant scatter in the residuals within both eclipses
(larger in the secondary), which we interpret as being caused by the
presence of spots on both stars. This is consistent with other
evidence of stellar activity such as the variability out of eclipse,
described later. This extra scatter, which represents correlated
(``red'') noise, raises the concern that it may be biasing the
solution, and causing the uncertainties to be underestimated.

To address this concern, we reanalyzed the data using the complete
light curve for \epic, with the instrumental effects removed but
without normalizing it by the spline fit indicated in
Section~\ref{sec:photometry}, i.e., preserving the out-of-eclipse
variability. We divided the data set into 25 segments, each containing
two primary and two secondary eclipses, and performed independent
solutions in each segment. To account for the distortions caused by
spots, we added a four-term Fourier series to the model (nine extra
parameters). The period was kept fixed at the value from
Table~\ref{tab:mcmc}, and the mass ratio was held at its value in
Table~\ref{tab:specorbit}. Reflection effects were modeled by solving
for the albedo coefficients for each component ($A_{\rm Aa}$, $A_{\rm
  Ab}$), with loose Gaussian priors set to average values of 0.5, as
appropriate for stars with convective envelopes, and a standard
deviation of 0.3.

%
\setlength{\tabcolsep}{6pt}
\begin{deluxetable}{lc}
\tablewidth{1.0\columnwidth}
\tablecaption{Results of Independent MCMC Analyses for 25 Segments of
  \epic\ Photometry, with the Addition of a 4-term Fourier Series to
  the Model\label{tab:cycles}}

\tablehead{ \colhead{~~~~~~~~~~~Parameter~~~~~~~~~~~} & \colhead{Value} }
\startdata
 $J$                              &  $0.693^{+0.020}_{-0.010}$                \\ [1ex]
 $r_{\rm Aa}+r_{\rm Ab}$          &  $0.25335^{+0.00044}_{-0.00100}$          \\ [1ex]
 $k$                              &  $0.942^{+0.036}_{-0.023}$                \\ [1ex]
 $\cos i$                         &  $0.0674^{+0.0018}_{-0.0032}$             \\ [1ex]
 $\sqrt{e}\cos\omega_{\rm Aa}$    &  $+0.0021^{+0.0050}_{-0.0056}$            \\ [1ex]
 $\sqrt{e}\sin\omega_{\rm Aa}$    &  $-0.074^{+0.015}_{-0.017}$               \\ [1ex]
 $A_{\rm Aa}$                     &  $0.422^{+0.084}_{-0.086}$                \\ [1ex]
 $A_{\rm Ab}$                     &  $0.26^{+0.18}_{-0.18}$                   \\ [1ex]
 $f_{K2}$                         &  $0.51^{+0.14}_{-0.12}$                   \\ [1ex]
\noalign{\hrule} \\ [-1.5ex]
\multicolumn{2}{c}{Derived quantities} \\ [0.5ex]
\noalign{\hrule} \\ [-1.5ex]
 $r_1$                            &  $0.1307^{+0.0017}_{-0.0022}$             \\ [1ex]
 $r_2$                            &  $0.1234^{+0.0029}_{-0.0013}$             \\ [1ex]
 $i$ (degree)                     &  $86.14^{+0.16}_{-0.11}$                  \\ [1ex]
 $e_{\rm A}$                      &  $0.0052^{+0.0047}_{-0.0027}$             \\ [1ex]
 $\omega_{\rm Aa}$ (degree)       &  $271.6^{+4.0}_{-3.5}$                    \\ [1ex]
 $\ell_{\rm Ab}/\ell_{\rm Aa}$    &  $0.595^{+0.039}_{-0.044}$                \\ [1ex]
 $J_{\rm ave}$                    &  $0.683^{+0.020}_{-0.009}$                
\enddata

\tablecomments{The values listed correspond to the median of the
  independent results for the 25 segments of the light curve, with the
  corresponding 68.3\% confidence levels.}

\end{deluxetable}
\setlength{\tabcolsep}{6pt}

The median value for each parameter over the 25 segments, and the
corresponding 68.3\% confidence intervals from the dispersion of the
25 solutions, are given in Table~\ref{tab:cycles}. The agreement with
the results in Table~\ref{tab:mcmc} is excellent, suggesting our
initial solution is unbiased.  The uncertainties, however, are
considerably larger. We have chosen to adopt these more conservative
errors here, and assign them to the parameter values in
Table~\ref{tab:mcmc}. We note that the eccentricity in
Table~\ref{tab:cycles} appears much less significant than before. It
is now different from zero at just under the 2$\sigma$ level.

\section{Absolute Dimensions}
\label{sec:dimensions}

We report the physical properties of the \epic\ components in
Table~\ref{tab:dimensions}. The masses have fractional errors of 1.4
and 1.2\% for the primary and secondary, and the radii are good to
about 1.8\% for both stars. The measured $v \sin i$ values are both
marginally larger than those predicted assuming spin-orbit
synchronization and alignment for a circular orbit ($v_{\rm sync}$ in
the table), although the difference may not be significant.

%
\begin{deluxetable}{lcc}
\tablewidth{1.0\columnwidth}
\tablecaption{Physical Properties of \epic \label{tab:dimensions}}
\tablehead{ \colhead{~~~~~~~~~~Parameter~~~~~~~~~~} & \colhead{Primary} & \colhead{Secondary} }
\startdata
 $M$ ($\mathcal{M}_{\sun}^{\rm N}$)         &  $0.912 \pm 0.013$             &  $0.822 \pm 0.010$    \\ [1ex]
 $R$ ($\mathcal{R}_{\sun}^{\rm N}$)         &  $0.920 \pm 0.016$             &  $0.851 \pm 0.016$    \\ [1ex]
 $\log g$ (dex)                             &  $4.470 \pm 0.016$             &  $4.494 \pm 0.017$    \\ [1ex]
 $q \equiv M_2/M_1$                         &          \multicolumn{2}{c}{$0.9003 \pm 0.0087$}       \\ [1ex]
 $a$ ($\mathcal{R}_{\sun}^{\rm N}$)         &          \multicolumn{2}{c}{$6.979 \pm 0.052$}         \\ [1ex]
 $T_{\rm eff}$ (K)                          &  $5035 \pm 150$                &  $4690 \pm 130$       \\ [1ex]
 $\Delta T_{\rm eff}$ (K)                   &          \multicolumn{2}{c}{$345 \pm 60$}              \\ [1ex]
 $L$ ($L_{\sun}$)                           &  $0.490 \pm 0.060$             &  $0.316 \pm 0.038$    \\ [1ex]
 $M_{\rm bol}$ (mag)                        &  $5.51 \pm 0.13$               &  $5.98 \pm 0.13$      \\ [1ex]
 $BC_V$ (mag)                               &  $-0.29 \pm 0.12$              &  $-0.47 \pm 0.13$     \\ [1ex]
 $M_V$ (mag)                                &  $5.80 \pm 0.22$               &  $6.46 \pm 0.24$      \\ [1ex]
 $v_{\rm sync} \sin i$ (\kms)\tablenotemark{a} &  $28.6 \pm 0.5$             &  $26.5 \pm 0.5$       \\ [1ex]
 $v \sin i$ (\kms)\tablenotemark{b}         &  $32 \pm 3$                    &  $31 \pm 4$           \\ [1ex]
 $E(B-V)$ (mag)                             &          \multicolumn{2}{c}{0.12~$\pm$~0.03}           \\ [1ex]
 $A_V$ (mag)                                &          \multicolumn{2}{c}{$0.372 \pm 0.093$}         \\ [1ex]
 Distance modulus (mag)                     &          \multicolumn{2}{c}{$7.32 \pm 0.23$}           \\ [1ex]
 Distance (pc)                              &          \multicolumn{2}{c}{$290 \pm 30$}              \\ [1ex]
 $\pi$ (mas)                                &          \multicolumn{2}{c}{$3.44 \pm 0.37$}           \\ [1ex]
 $\pi_{Gaia/{\rm EDR3}}$ (mas)\tablenotemark{c} &      \multicolumn{2}{c}{$3.479 \pm 0.022$}         
\enddata

\tablecomments{The masses, radii, and semimajor axis $a$ are expressed
  in units of the nominal solar mass and radius
  ($\mathcal{M}_{\sun}^{\rm N}$, $\mathcal{R}_{\sun}^{\rm N}$) as
  recommended by 2015 IAU Resolution B3 \citep[see][]{Prsa:2016}, and
  the adopted solar temperature is 5772~K (2015 IAU Resolution
  B2). Bolometric corrections are from the work of \cite{Flower:1996},
  with conservative uncertainties of 0.1~mag, and the bolometric
  magnitude adopted for the Sun appropriate for this $BC_V$ scale is
  $M_{\rm bol}^{\sun} = 4.732$ \citep[see][]{Torres:2010}. See text
  for the source of the reddening. For the apparent visual magnitude
  of \epic\ out of eclipse we used $V = 13.013 \pm 0.029$
  \citep{Henden:2015}. }

\tablenotetext{a}{Synchronous projected rotational velocity assuming a
  circular orbit and spin-orbit alignment.}

\tablenotetext{b}{Measured projected rotational velocities.}

\tablenotetext{c}{A parallax zero-point correction of $+0.022$~mas has
  been added to the catalog parallax \citep{Lindegren:2021}.}

\end{deluxetable}

The spectroscopic temperatures in Section~\ref{sec:spectroscopy} are
not precise enough for a useful comparison with models. In particular,
the temperature difference is essentially unconstrained due to the
large errors. A much more precise and accurate measure of the
temperature ratio (or difference) can be obtained directly from the
light curve analysis, through the surface brightness ratio $J$ (or the
disk-integrated value $J_{\rm ave}$; see Table~\ref{tab:cycles}).
Additionally, a luminosity-weighted mean temperature can be derived
from available brightness measurements for \epic.  These two
quantities together allow us to infer the individual $T_{\rm eff}$
values.

Brightness measurements were gathered from the literature in the
Johnson, Sloan, and 2MASS photometric systems \citep{Henden:2015,
  Skrutskie:2006}. We constructed nine non-independent color indices,
and used color-temperature calibrations from \cite{Casagrande:2010}
and \cite{Huang:2015} to infer the mean temperature, setting the
metallicity to the known value for the parent cluster. We adopted a
reddening of $E(B-V) = 0.12 \pm 0.03$, which is the average of our
independent determinations for three other eclipsing binaries in
\rup\ (Papers~I, II, and III). A systematic difference of about 130~K
exists between the temperatures inferred from the above two
calibrations \citep[see][]{Huang:2015}. For consistency with our
earlier studies, we adjusted the results from the Casagrande
calibration to place them on the Huang zero point, and then averaged
all individual values. The result is a mean system temperature of
$4900 \pm 100$~K.

The temperature difference we obtained from the disk-integrated
surface brightness ratio $J_{\rm ave}$ is $345 \pm 60$~K. The mean
system temperature then leads to individual values of $5035 \pm 150$~K
for the primary and $4690 \pm 130$~K for the secondary. They
correspond to spectral types of K2 and K4. We list these results in
Table~\ref{tab:dimensions}.

The luminosities together with bolometric corrections from
\cite{Flower:1996} \citep[see also][]{Torres:2010} and an adopted
visual magnitude of $V = 13.013 \pm 0.029$ \citep{Henden:2015} lead to
a distance of $290 \pm 30$~pc, corresponding to a parallax of $3.44
\pm 0.37$~mas. This is in excellent agreement with the parallax listed
in the Gaia EDR3 catalog \citep[$3.479 \pm 0.022$~mas, after
  corrections for a zero-point bias following][]{Lindegren:2021}, and
supports the accuracy of our radius and temperature determinations.

With our mass estimates for the binary components and the
spectroscopic solution for the outer orbit of the triple system, we
calculate a minimum mass for the tertiary of 0.27~$M_{\sun}$ for an
edge-on orbit. If it is a main-sequence star, the upper limit on its
brightness, transformed to the $K\!p$ band, corresponds to
approximately 2\% of the flux of the primary star.  Appealing to an
isochrone for the cluster from the PARSEC~1.2S series of
\cite{Chen:2014}, with ${\rm [Fe/H]} = +0.10$ and an age of 2.67~Gyr
as determined from Papers I--III, this relative flux leads to a
maximum tertiary mass of about 0.47~$M_{\sun}$. This places a lower
limit on the inclination angle of the outer orbit of
$\sim$38\arcdeg\ relative to the line of sight. However, it is also
possible that the tertiary is a more massive white dwarf. This
alternative scenario will be discussed below.

\section{Rotation and Activity}
\label{sec:rotation}

\begin{figure}
\epsscale{1.15}
\plotone{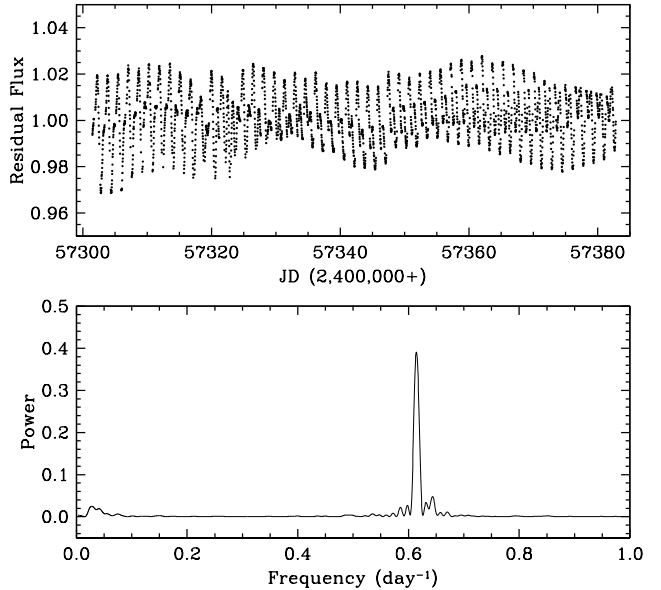}

\figcaption{\emph{Top:} Out-of-eclipse variability in the K2
  photometry of \epic, interpreted here as due to rotational
  modulation by spots on one or both stars. The binary eclipses have
  been masked out. \emph{Bottom:} Lomb-Scargle periodogram featuring a
  peak at a frequency corresponding to a rotational period of
  $1.628^{+0.014}_{-0.017}$~days, similar to the orbital
  period.\label{fig:rotation}}

\end{figure}

The light curve of \epic\ displays obvious variability out of eclipse
that we interpret as a signature of rotation
(Figure~\ref{fig:rotation}, top panel). The peak-to-peak amplitude
over the entire 81-day time series is almost 6\%, indicating a
significant level of activity. The Lomb-Scargle periodogram in the
lower panel displays a main peak at a frequency corresponding to
$P_{\rm rot} = 1.628^{+0.014}_{-0.017}$~days, where the uncertainty
was calculated from the half width of the peak at half maximum. This
period is consistent with the orbital period, within the
uncertainties, suggesting spin-orbit synchronization.

\begin{figure*}
  \includegraphics[height=5.5cm]{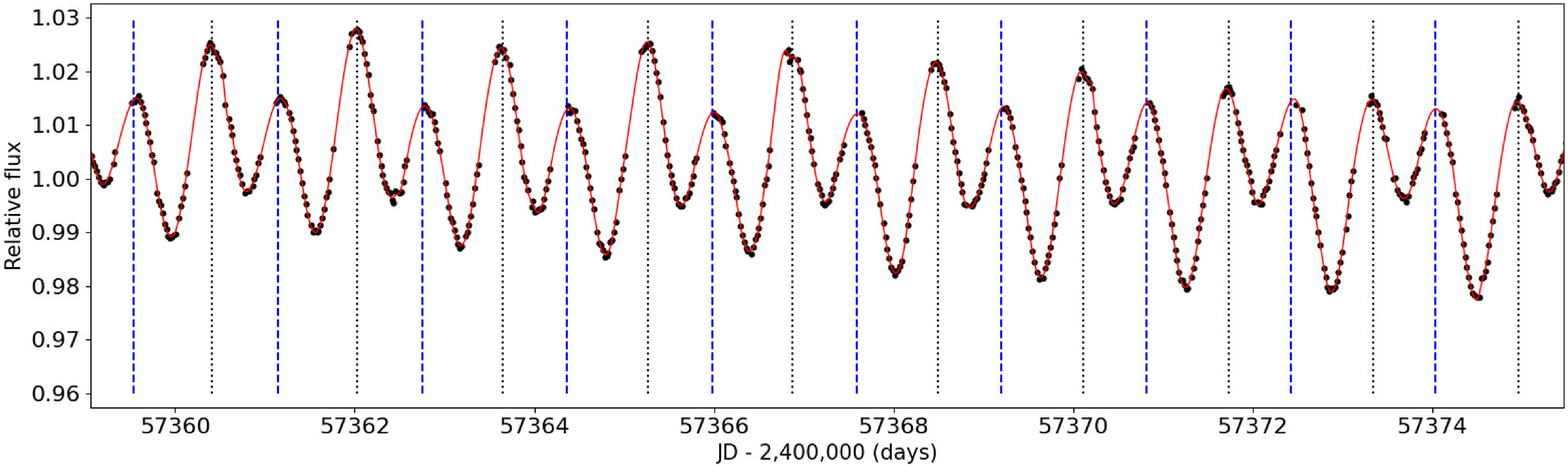}
  \figcaption{Section of the light curve of \epic\ showing the
    presence of two signals with slightly different periods. The red
    line is a spline fit to the data to guide the eye.  Alternating
    peaks are marked with dotted and dashed lines, and have apparent
    periods of 1.617 and 1.612 days, respectively.\label{fig:rotzoom}}
\end{figure*}

The rotational modulation appears quite complex. Closer inspection
reveals the presence of two distinct signals with slightly different
periods.  The two signals are marked with dotted and dashed lines in
Figure~\ref{fig:rotzoom}, which is an enlargement of a section of the
top panel of the previous figure. For each signal the intervals
between consecutive maxima in the light curve were determined by eye,
and are not quite the same: one signal (dotted lines) has a
periodicity of 1.617 days, while the other (dashed) repeats every
1.612 days. We estimate the uncertainty in each of these estimates to
be about 0.002 days.  Both periods are formally consistent with the
one measured from the periodogram in Figure~\ref{fig:rotation}, which
has a larger uncertainty, and they both appear to be marginally
shorter than the orbital period of 1.622~days. The individual
amplitudes of the signals are difficult to pin down, and seem to vary
with time.

It is unclear whether the signals correspond to different components,
or to only one of the stars. The latter scenario could result from two
spots or spot regions at different latitudes (and longitudes) on the
same star, carried around at slightly different speeds due to
differential rotation. The primary star is about 40\% brighter, but
the secondary could well be more active, so the relative strengths of
the signals may actually be similar. It seems likely, therefore, that
both components are contributing to the variability out-of-eclipse.

\begin{figure}
\epsscale{1.15}
\plotone{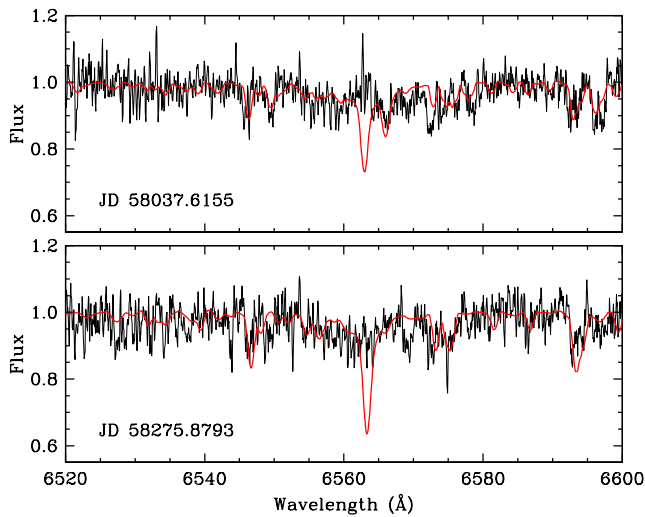}

\figcaption{Two of our spectra for \epic\ in the H$\alpha$ region,
  compared against synthetic double-lined spectra (red lines)
  constructed from individual PHOENIX models for each component,
  Doppler shifted according to their measured velocities, and scaled
  in accordance with their flux ratio. Julian dates ($-$2,400,000) for
  the observations are indicated in each panel. \label{fig:halpha}}
\end{figure}

Other common signatures of activity include the presence of emission
cores in the \ion{Ca}{2}~H and K lines, and the H$\alpha$ line in
emission. The flux in the blue part of our spectra is too low to
detect the calcium lines, but the H$\alpha$ feature does suggest a
modest level of activity.  Figure~\ref{fig:halpha} shows two of our
best spectra in this region at different orbital phases, compared
against double-lined synthetic spectra for the appropriate phases. We
generated the synthetic spectra using PHOENIX models from
\cite{Husser:2013}, for the temperatures and surface gravities of the
two components and an estimated flux ratio of 0.64 at this wavelength.
The H$\alpha$ line is missing, or perhaps slightly in emission in the
top panel. At the very least the line appears to be completely filled
in, which is evidence of chromospheric activity.

Finally, \epic\ has been detected in X-rays by the Swift mission
\citep{Evans:2020}, in observations carried out between 2014 and 2018.
This, and the other signs of activity, are driven by the tidally
induced rapid rotation of the stars, which is more than an order of
magnitude faster than single stars of this mass in the cluster
\citep[see][]{Curtis:2020}.

\section{Comparison with Models}
\label{sec:models}

Precisely measured masses, radii, and effective temperatures of stars
in eclipsing binaries can provide important tests of stellar evolution
theory, particularly when the masses and radii are determined to
better than 3\%, as is the case for \epic\ \citep[see,
  e.g.,][]{Andersen:1991, Torresetal:2010}. In many of these
comparisons with models the metallicity of the system and its age are
often not known, and must be left as free parameters. This tends to
weaken the test. Membership of \epic\ in \rup\ makes it an ideal case
for such a comparison, because estimates are available for both of
those properties.  The metallicity is close to ${\rm [Fe/H]} = +0.10$
\citep{Curtis:2018, Bragaglia:2018}, and our earlier studies in
Papers~I--III of three other eclipsing binaries in \rup\ have provided
an age estimate for the cluster of $2.67 \pm 0.21$~Gyr, based on the
PARSEC~1.2S models of \cite{Chen:2014}.

\begin{figure}
\epsscale{1.15}
\plotone{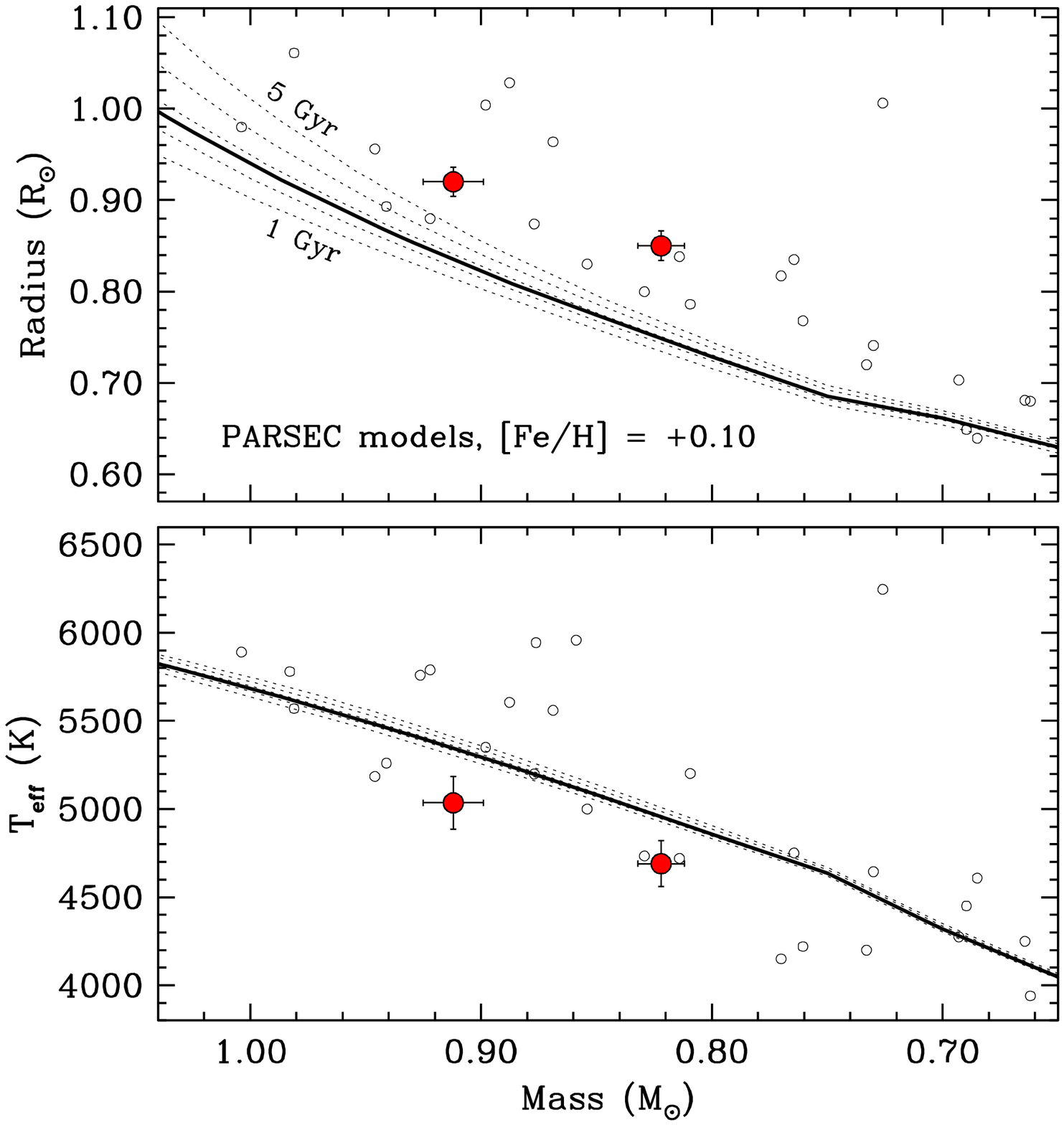}

\figcaption{Comparison of the mass, radius, and effective temperature
  measurements for \epic\ against isochrones from the PARSEC~1.2S
  series of \cite{Chen:2014}, for the adopted metallicity of the
  \rup\ cluster, ${\rm [Fe/H]} = +0.10$. The heavy line corresponds to
  the age of 2.67~Gyr determined from three previously studied
  eclipsing binaries in the cluster \citep{Torres:2018, Torres:2019,
    Torres:2020}. Dotted lines represent isochrones from 1 to 5~Gyr
  from the same models. Masses, radii, and temperatures for other
  eclipsing binaries from the literature that are in the same mass
  regime as \epic\ and have fractional errors below 3\% are shown for
  context. \label{fig:parsec}}
\end{figure}

Figure~\ref{fig:parsec} compares the measured properties of the
\epic\ components with the same models as above, for the known
metallicity and age of the parent cluster. Other eclipsing binaries
from the literature with components that have masses and radii
measured to 3\% or better and are in the same mass range are shown as
well.  At the measured masses for \epic, the radii of both components
are significantly larger than predicted, by about 10\% (5.3$\sigma$)
for the primary and 14\% (6.4$\sigma$) for the secondary. At the same
time, the effective temperatures are seen to be cooler than predicted,
by 310~K and 270~K, respectively, or roughly 6\%
(2$\sigma$).\footnote{Note that the primary and secondary temperatures
  are strongly correlated, and that their difference is much better
  determined than their absolute values through the use of $J_{\rm
    ave}$, as described earlier. The models predict a
  primary/secondary difference of 390~K, quite close to the measured
  value of $345 \pm 60$~K.}  Both of these trends go in the same
direction as observed in many other active lower main-sequence stars
\citep[e.g.,][]{Torres:2013}, and point to the activity of \epic\ as
the cause of the deviations from theory. The systematic deviations are
usually referred to as ``radius inflation'' and ``temperature
suppression''.

On the other hand, the bolometric luminosities predicted by the
models, $L_{\rm Aa} = 0.51~L_{\sun}$ and $L_{\rm Ab} = 0.31~L_{\sun}$,
are in good agreement with our empirical estimates of $0.490 \pm
0.060$ and $0.316 \pm 0.038~L_{\sun}$ for the primary and secondary,
respectively. This is a direct result of the Stefan-Boltzmann law, and
the fact that the fractional radius discrepancies with the models are
roughly twice as large as those in the temperatures.

It is worth noting that the PARSEC models we have used for this
comparison differ from other standard models in that they have been
modified by their builders in an attempt to ``correct'' the problems
of radius inflation and temperature suppression. As explained by
\cite{Chen:2014}, this was done by adjusting the temperature-opacity
relation for stars less massive than about 0.7~$M_{\sun}$ (at solar
metallicity), in such a way as to increase the radii enough to match
observations of low-mass stars in the mass-radius diagram. Both
components of \epic\ are more massive than 0.7~$M_{\sun}$, so those
adjustments should not have an impact on them, and the PARSEC models
effectively behave as standard models for these stars.

\section{Discussion and Final Remarks}
\label{sec:conclusions}

It was our expectation at the beginning of this study that precise and
accurate masses, radii, and effective temperatures for \epic\ would
allow for a stringent test of stellar evolution models and an
independent estimate of the age of \rup, to supplement similar
determinations for three other eclipsing binaries in the cluster
analyzed previously. While the significant level of activity in
\epic\ has prevented us from doing that, it has rewarded us with a
rare opportunity for an assumption-free estimate of the degree of
radius inflation and temperature suppression, because of the fact that
the metallicity and age of \rup\ are already known. This is not
usually the case for a typical low-mass eclipsing binary in the
field. The radius and temperature deviations from theory in those
cases are often measured with reference to an arbitrary isochrone of
solar metallicity and some fixed age that varies from author to
author, so they depend to some extent on those assumptions.  Age may
well have a relatively small effect, especially for M dwarfs, but that
is not the case for the metallicity \citep[see, e.g.,][]{Berger:2006}.

Stellar activity such as we see in \epic\ is typically associated with
strong magnetic fields and/or the presence of spots.  Magnetic fields
tend to inhibit convection and slow the outward transport of energy
\cite[e.g.,][]{Chabrier:2007}, and the stars adjust by increasing
their surface area and reducing their surface temperature. Star spots
are thought to elicit a similar response \citep{Chabrier:2007,
  Somers:2015}.  Although most publicly available ``standard'' models
of stellar evolution do not account for these effects, a few efforts
in this direction to include non-standard physics have been made, with
promising results \citep[e.g.,][]{DAntona:2000, MacDonald:2009,
  MacDonald:2014, Feiden:2013, Feiden:2014, Somers:2015,
  Somers:2020}. Of the small number of detailed comparisons of
eclipsing binary observations against these non-standard models that
have been made, almost all have involved active M dwarfs. One
exception is the F+K system EF~Aqr \citep{Feiden:2012}.
\epic\ therefore represents an ideal case to test those models for a
pair of K dwarfs with absolute masses, radii, and temperatures
measured to better than 3\%, and a known age and metallicity.

Turning now to the multiplicity of \epic, it is interesting to note
that this is the second hierarchical triple system found among the
four eclipsing binaries in \rup\ that we have studied thus far. This
is not surprising given its 1.6~day orbital period, as
\cite{Tokovinin:2006} have shown that the vast majority ($\sim$96\%)
of spectroscopic binaries with periods under 3 days have additional
components. The other triple system we have found is
\epicthird\ \citep{Torres:2020}, with a period of
2.75~days.\footnote{The remaining two eclipsing binaries,
  \epicfirst\ and \epicsecond\ \citep{Torres:2018, Torres:2019}, have
  longer periods of 6.53 and 11.99~days, and are not known to have
  third components.}

As was the case in that system, the nature of the third star in the
present one is unclear.  If it is a lower main-sequence star, we have
estimated its mass to be in the range 0.27--0.47~$M_{\sun}$,
corresponding to a mid-to-late M dwarf. If a white dwarf, it could be
more massive.  In either case, the third star may affect the dynamical
evolution of the system. For example, it could modulate the angle
between the inner and outer orbital planes through Kozai-Lidov
oscillations \citep[e.g.,][]{Naoz:2016}, as well as the eccentricity
of the inner eclipsing system. We have, in fact, found some evidence,
tentative as it may be, that the inner orbit may be slightly
eccentric, whereas given the present age of the system it would be
expected to have long been circularized \citep[on a timescale of
  $\sim$10~Myr; e.g.,][]{Hilditch:2001}. Confirmation of this non-zero
eccentricity must await new observations, preferably high-precision
light curves that are more sensitive to the eccentricity. Current
observing plans for NASA's TESS mission do not include pointings near
\epic\ for the near future. Based on the dynamical stability criteria
of \cite{Eggleton:1995} and \cite{Mardling:2001}, we find that for any
reasonable mass of the third star and any relative inclination of the
orbital planes, the \epic\ system is in no danger of being disrupted.

The angular size of the outer orbit is estimated to be about 3~mas.
While this would be within reach of long-baseline interferometers such
as CHARA, VLTI, or NPOI, the system is likely too faint for the
sensitivity of current instrumentation ($V = 13.0$, $K = 10.6$).  The
Gaia mission cannot spatially resolve the third star, but could in
principle detect the wobble of the center of light of the system. The
renormalized unit weight error (RUWE) from the Gaia EDR3 catalog,
which is indicative of the quality of the astrometric solution, has a
value for \epic\ of 1.183. This is nominally within the range of
1.0--1.4 in which the solutions are considered acceptable\footnote{See
  report by L.\ Lindegren in the Gaia documentation,
  \url{https://www.cosmos.esa.int/web/gaia/public-dpac-documents}},
although \cite{Stassun:2021} have reported based on a large sample of
eclipsing binaries that even within that range the RUWE values tend to
correlate with the semimajor axis of the motion of the
photocenter. Another useful indicator of a perturbed astrometric
solution is the {\tt astrometric\_excess\_noise\/} parameter, for
which Gaia gives the value 0.09~mas. The statistical significance
attached to this quantity is $D = 5.23$ (a dimensionless measure of
significance). Values larger than about $D = 2.0$ are regarded as a
sign that there may be unmodeled motion in the source.  It is
possible, therefore, that Gaia has already detected the signature of
the triple system. If confirmed, this could eventually yield the
inclination angle of the outer orbit, which would then enable the mass
of the tertiary to be determined, perhaps clarifying its true nature.


\begin{acknowledgements}

The spectroscopic observations of \epic\ were gathered with the help
of P.\ Berlind, M.\ Calkins, and G.\ Esquerdo. J.\ Mink is thanked for
maintaining the CfA echelle database. We thank the anonymous
  referee for helpful comments. G.T.\ acknowledges partial support
from NASA's Astrophysics Data Analysis Program through grant
80NSSC18K0413. J.L.C.\ is supported by the NSF Astronomy and
Astrophysics Postdoctoral Fellowship under award AST-1602662, and by
NASA under grant NNX16AE64G issued through the K2 Guest Observer
Program (GO~7035).  E.G.\ acknowledges support from NSF Astronomy \&
Astrophysics grant 1817215.
The research has
made use of the SIMBAD and VizieR databases, operated at the CDS,
Strasbourg, France, and of NASA's Astrophysics Data System Abstract
Service.
The work has also made use of data from the European Space Agency
(ESA) mission Gaia (\url{https://www.cosmos.esa.int/gaia}), processed
by the Gaia Data Processing and Analysis Consortium (DPAC,
\url{https://www.cosmos.esa.int/web/gaia/dpac/consortium}). Funding
for the DPAC has been provided by national institutions, in particular
the institutions participating in the Gaia Multilateral Agreement. The
computational resources used for this research include the Smithsonian
Institution's ``Hydra'' High Performance Cluster.

\end{acknowledgements}

\end{document}